\documentclass[]{emulateapj}
\usepackage{placeins}
\usepackage{graphicx}
\usepackage[normalem]{ulem}
\usepackage{amssymb, amsmath}
\newcommand{\rns}{\rho_{\rm sat}}
\newcommand{\Ms}{M_{\odot}}

\begin{document}

\title{From Neutron Star Observables to the Equation of State.\\  I. An Optimal Parametrization}
\author{Carolyn A. Raithel, Feryal \"Ozel, \& Dimitrios Psaltis}
\affiliation{Department of Astronomy and Steward Observatory, University of Arizona, 933 N. Cherry Avenue, Tucson, Arizona 85721, USA}

\begin{abstract}
The increasing number and precision of measurements of neutron star masses, radii, and, in the near future, moments of inertia offer the possibility of precisely determining the neutron star equation of state. One way to facilitate the mapping of observables to the equation of state is through a parametrization of the latter. We present here a generic method for optimizing the parametrization of any physically allowed EoS. We use mock equations of state that incorporate physically diverse and extreme behavior to test how well our parametrization reproduces the global properties of the stars, by minimizing the errors in the observables mass, radius, and the moment of inertia. We find that using piecewise polytropes and sampling the EoS with five fiducial densities between $\sim1-8$ times the nuclear saturation density results in optimal errors for the smallest number of parameters. Specifically, it recreates the radii of the assumed EoS to within less than 0.5 km for the extreme mock equations of state and to within less than 0.12 km for 95\% of a sample of 42 proposed, physically-motivated equations of state. Such a parametrization is also able to reproduce the maximum mass to within 0.04~$\Ms$ and the moment of inertia of a 1.338~$\Ms$ neutron star to within less than 10\% for 95\% of the proposed sample of equations of state.
\end{abstract}

\maketitle

\section{Introduction}
A key goal of observations of neutron stars is to determine the equation of state of cold, ultradense matter. The equation of state at high densities results from the interactions of nucleons, quarks, and possibly other constituents and fully characterizes the microphysics of the neutron star interior. Over the years, the sample of proposed equations of state (EoS) have incorporated vastly different physics and, accordingly, have predicted a variety of different global properties for neutron stars (see \citealt{Ozel2016} for a recent review). Some early studies employed a purely nucleonic framework (e.g., \citealt{Baym1971, Friedman1981, Akmal1998, Douchin2001}), while others explored the role of hyperons (e.g., \citealt{Balberg1997}), pion condensates (e.g., \citealt{Pandharipande1975}), and kaon condensates (e.g., \citealt{Kaplan1986}). It is even possible, albeit contrived, to construct EoS that generate parallel stable branches of neutron stars with the same masses but different radii by introducing a first-order phase transition into the EoS of hybrid neutron stars (\citealt{Glendenning2000, Blaschke2013}). More recently, there have been studies that incorporate the expected quark degrees of freedom at high densities, either from phenomenological models or deriving from early results of lattice QCD (e.g., \citealt{Alford2005, Alford2013, Kojo2015}). 

Nuclear physics experiments are able to constrain these various EoS only up to densities near the nuclear saturation density, $\rns \sim 2.7\times10^{14}$~g~cm$^{-3}$ (see \citealt{Lattimer2012}). In this regime, there have been attempts to convert experimental results more directly into physical constraints. Even though, at these densities, the interactions between particles can be formally described in terms of static, few-body potentials (see, e.g., \citealt{Akmal1998, Morales2002, Gandolfi2012}), there are large uncertainties arising from extrapolations to $\beta$-equilibrium and low-temperature matter. Furthermore, at densities well above $\rns$, the interactions between particles can no longer be expanded in these terms.

At densities beyond $\rns$, experimentally constraining the dense-matter EoS can be accomplished using observations of neutron stars and exploiting the direct mapping between the EoS and the mass-radius relation \citep{Lindblom1992}.  This can be achieved by comparing measured masses (\citealt{Demorest2010, Antoniadis2013}) and radii (\citealt{Guillot2013, Guillot2014, Heinke2014, Nattila2015, Ozel2016a, Bogdanov2016}; see \citealt{Ozel2016} for a recent review) to the predictions of particular EoS. However, this approach is limited in scope, as it serves to constrain the parameters of already formulated EoS frameworks and may not necessarily span the full range of possibilities. In other words, it may not be capable of recreating the EoS from observations in a model-independent way.

A separate approach is to empirically infer the EoS from the combination of these radii and mass measurements by utilizing a model-independent parametrization of the EoS. To date, several parametrizations have been proposed. The EoS can be written as a spectral expansion in terms of the enthalpy  \citep{Lindblom2012, Lindblom2014}. Alternatively, the EoS can be represented as a discrete number of segments that are piecewise polytropic or linear (\citealt{Read2009, Ozel2009, Hebeler2010, Steiner2016}).

Despite these various proposals, only limited systematic optimizations have been undertaken to determine the ideal number of segments and functional forms for a parametric EoS. Moreover, these earlier studies used an existing sample of theoretical EoS as benchmarks and sought to represent only this subset. Finally, many of these early studies only sought to reproduce the predicted radii to rather large uncertainties by today's standards, reflecting the available data at the time. However, the current sample of mass and radius measurements, as well as the anticipated moment of inertia measurement from the binary system PSR J0737$-$3039, which is expected to be measured with 10\% accuracy in the next five years \citep{Lyne2004, Kramer2009}, require reconstructions of the EoS to a much higher accuracy than ever before. Any useful parametrization must be able to recreate masses and radii to high precision with a minimum set of parameters.

We present here an optimization of a piecewise-polytropic parametric EoS. In order to test our parametrization as generally as possible, we generate mock EoS that incorporate physically extreme behavior. We then apply our parametrization to these mock EoS and determine the differences in masses, radii, and moments of inertia between the parametrized and the full EoS. We optimize the number of segments included in the parametric EoS by requiring these differences to be smaller than the expected accuracy of observations. We find that sampling the EoS with five fiducial densities that are evenly spaced in the logarithm of density (i.e., using five polytropes) recreates the radii of the assumed EoS to within $\lesssim$0.5 km for the extreme cases and to within $\lesssim$0.12 km for 95\% of a sample of 42 proposed EoS with a wide range of input physics. Such a parametrization is also able to reproduce the maximum mass to within $\lesssim$~0.04$\Ms$ and the moment of inertia to within $\lesssim$10\% for 95\% of the proposed EoS sample.

\section{Optimizing the Parametric EoS} 
\label{sec:optimizing}
\begin{figure}[ht]
\centering
\includegraphics[width=0.4\textwidth]{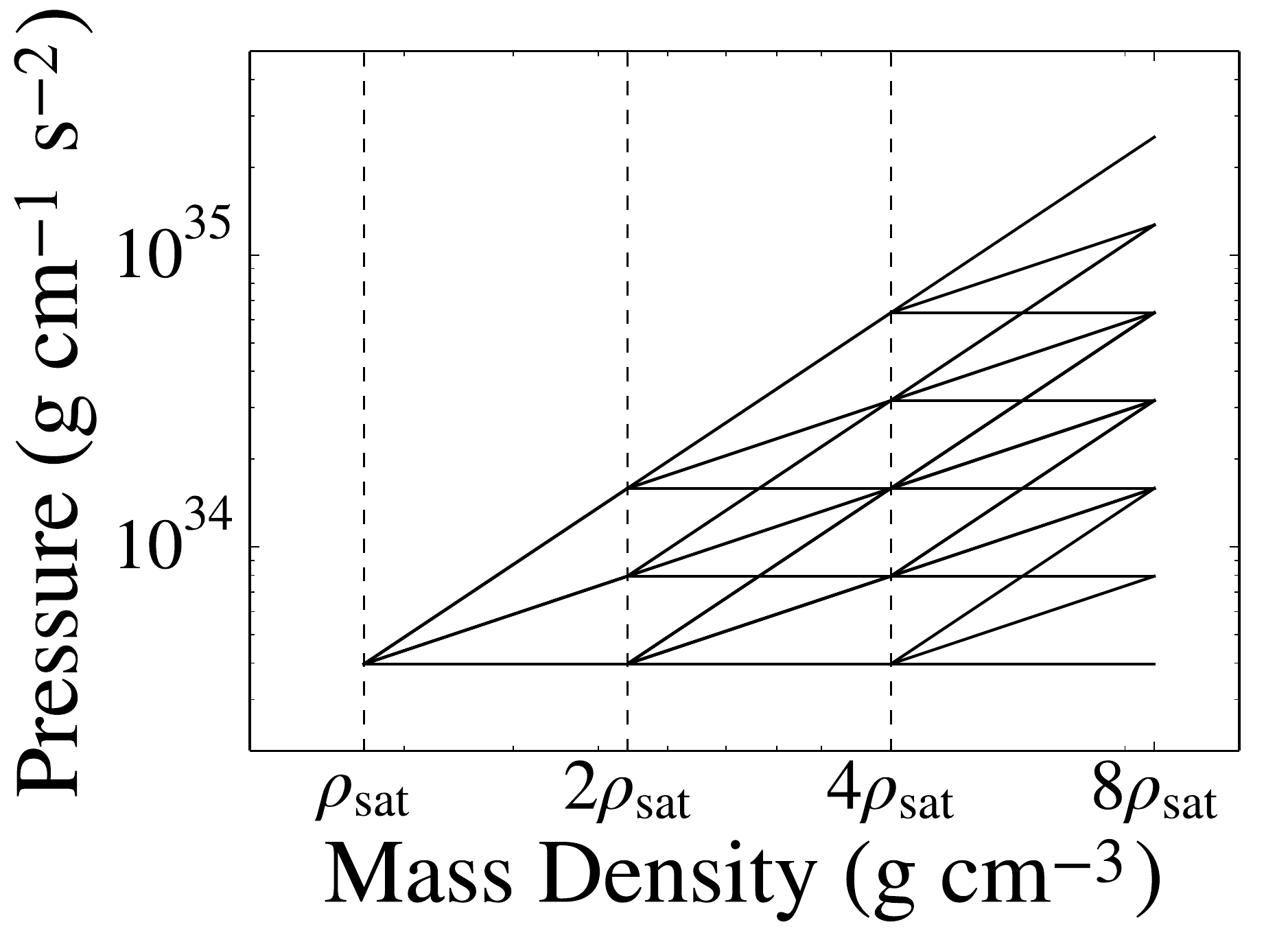}
\caption{\label{fig:polytrope}  Pressure as a function of density for a sample of piecewise polytropes. The equation of state is divided into polytropic segments at three fiducial densities that are uniformly spaced in the logarithm. In each segment, we allow the polytrope in equation~(\ref{eq:polytrope}) to have an index of $\Gamma$~=~0, 1, or 2 to illustrate their general behavior.}
\end{figure}

We parametrize the EoS in terms of $n$ piecewise polytropes,\footnote{See the Appendix for a discussion of a linearly parametrized EoS.} spaced between two densities, $\rho_{\rm min}$ and $\rho_{\rm max}$. We define the dividing density and pressure between each piecewise polytrope to be $\rho_i$ and $P_i$, respectively. Each polytropic segment is then given by
\begin{equation}
\label{eq:polytrope}
P = K_i \rho ^{\Gamma_i} \quad (\rho_{i-1} \le \rho \le \rho_i) ,
\end{equation}
where the constant, $K_i$, is determined from the pressure and density at the previous fiducial point according to
\begin{equation}
\label{eq:ki}
K_i = \frac{P_{i-1}}{\rho_{i-1}^{\Gamma_i}} = \frac{P_i}{\rho_i^{\Gamma_i}}
\end{equation}
and the polytropic index for the segment, $\Gamma_i$, is given by
\begin{equation}
\label{eq:gammai}
\Gamma_i = \frac{\log_{10}(P_i/P_{i-1})}{\log_{10}(\rho_i/\rho_{i-1})}.
\end{equation}
Figure \ref{fig:polytrope} shows an example of piecewise polytropes over three density segments, with various values of their polytropic indices, $\Gamma$, to illustrate the behavior of equations~(\ref{eq:polytrope})-(\ref{eq:gammai}).
Our primary goal in optimizing the parametrization is to reduce the errors in the prediction of observables (i.e., mass, radius, and moment of inertia) below a threshold that is comparable to the uncertainties in present or upcoming observations, while keeping the number of polytropic segments to a minimum. For the optimization process, we produce extreme, albeit physically allowed EoS between $1-8~\rns$ to test how well our parametrizations reproduce observables with various numbers of polytropic segments included. Once our parametrization is optimized, we then apply it to more reasonable, physically motivated EoS to test its ability to recreate those as well.

In addition to the number of polytropes to include in the parametrization, there are two other variables that we have to optimize: the density at which the parametrization should start and the spacing of the polytropic segments. For the question of where to start the parametrization, we explored starting at $\rho_0 = 10^{14}~$g~cm$^{-3}$ as well as at $\rho_0 = \rns$.  It is typically assumed that the EoS is known up to $\rns$; however, \citet{Lattimer2001} showed that for a sample of around 30 proposed EoS, the predicted pressures vary by a factor of 5 over the range $0.5~\rns < \rho < \rns$, even though these EoS are all meant to be consistent with nuclear physics experiments in this density regime. This is because the extrapolation of pressures from symmetric nuclear matter to neutron-rich matter is poorly constrained. Meanwhile, densities below $\sim0.5\rns$ do not significantly affect the global properties of the star. Therefore, we allowed our parametrization to start at both $0.5\rns$ and $\rns$, in order to explore these two limits. As for the question of how to space the polytropic segments, the preferred option is to space the segments evenly in the logarithm of the density. A logarithmic spacing more finely samples the low density region of the EoS, which is the region that most affects the resulting neutron star observables \citep{Lattimer2001, Read2009, Ozel2009}. For completeness, we also explored a second possibility: spacing the fiducial densities between each polytropic segment linearly.

We found that the combination of starting the parametrization at $\rns$ and spacing the fiducial densities evenly in the logarithm resulted in the smallest errors in mass and radius. We therefore start the first polytrope at $\rho_0 = \rns$ and space the remaining fiducial densities evenly in the logarithm between $\rho_0$ and 7.4~$\rns$. We set the last point, $\rho_n = 7.4~\rns$, following the results of \cite{Read2009} and \cite{Ozel2009} who found that the pressure at this density determines the neutron star maximum mass and that pressures at higher densities do not significantly affect the overall shape of the resulting mass-radius curve. We determined the pressure corresponding to each fiducial density by sampling whichever EoS we were parametrizing, i.e., $P_i = P_{\rm EoS}(\rho_i)$. For $\rho \le \rho_0$, we connected our parametrization to a low-density EoS. 

We varied the total number of fiducial densities above $\rns$ from 3 to 12. Clearly, as the number of polytropes used to represent the EoS increases, the errors in the observables are expected to reduce. Our goal in the remainder of this paper is to determine the minimum number of fiducial densities required to reproduce the mass, radius, and moment of inertia of a neutron star to within desired observational uncertainties.

\section{From EoS to Observables}
\label{sec:tov}
In order to determine how well our parametrizations were able to reproduce the observations predicted by a given EoS, we used the Tolman-Oppenheimer-Volkoff (TOV) equations and solved them to find the mass, radius, and moment of inertia. 

The TOV equations give the pressure, $P$, and the enclosed mass, $M$, of the star as a function of radius, according to
\begin{equation}
\frac{\text{d}P}{\text{d}r} = - \frac{G}{c^2} \frac{(\epsilon + P)(M + 4\pi r^3 P/c^2)}{r^2 - 2GMr/c^2}
\label{eq:dpdr}
\end{equation}
and
\begin{equation}
\frac{\text{d}M}{\text{d}r} = \frac{4 \pi r^2 \epsilon}{c^2},
\label{eq:dmdr}
\end{equation}
where the energy density, $\epsilon$, is given by
\begin{equation}
\label{eq:thermo}
d\frac{\epsilon}{\rho} = -P d\frac{1}{\rho}.
\end{equation}

To get the full relation between energy density and mass density, we can integrate equation~(\ref{eq:thermo}) for $\Gamma \ne 1$ to 
\begin{equation}
\epsilon(\rho) = (1+a)\rho c^2 + \frac{K}{\Gamma -1}\rho^{\Gamma},
\label{eq:eps_generic}
\end{equation}
where $a$ is an integration constant. Along any density section of the EoS, requiring continuity at either endpoint determines $a$ such that equation~(\ref{eq:eps_generic}) becomes
\begin{multline}
\label{eq:gammaNot1}
\epsilon(\rho) = \left[\frac{\epsilon(\rho_{i-1})}{\rho_{i-1}} - \frac{P_{i-1}}{\rho_{i-1}(\Gamma_i-1)}\right]\rho + \frac{K_i}{\Gamma_i -1} \rho^{\Gamma_i}, \\
 (\rho_{i-1} \le \rho \le \rho_i) 
\end{multline}
where $K_i$ and $\Gamma_i$ are determined as in equations~(\ref{eq:ki}) and (\ref{eq:gammai}). 

Similarly, for the case of $\Gamma$=1, equation~(\ref{eq:thermo}) becomes
\begin{multline}
\label{eq:gamma1}
\epsilon(\rho) = \frac{\epsilon(\rho_{i-1})}{\rho_{i-1}}\rho + K_i \ln\left(\frac{1}{\rho_{i-1}}\right)\rho - K_i \ln\left(\frac{1}{\rho}\right) \rho  \\
 \rho_{i-1} \le \rho \le \rho_i.
\end{multline}

We used equations~(\ref{eq:gammaNot1}) or (\ref{eq:gamma1}) to relate an EoS to the energy density, and then used that energy density to integrate the TOV equations outwards from the center of the star. The radius at which the pressure becomes negligible gives the total mass and radius of the star. 

In order to calculate the moment of inertia, we simultaneously solved equations~(\ref{eq:dpdr}) and (\ref{eq:dmdr}) with two coupled differential equations for the relativistic moment of inertia, 
\begin{equation}
\frac{\text{d}I}{\text{d}r} = \frac{8 \pi}{3}\frac{ \left( \epsilon + P \right)}{c^2} \frac{f j r^4}{1-2\frac{G M}{r c^2}},
\label{eq:didr}
\end{equation}
and
\begin{equation}
\frac{\text{d}}{\text{d}r} \left(r^4 j \frac{\text{d}f}{\text{d}r} \right) + 4 r^3 \frac{\text{d}j}{\text{d}r} f = 0,
\label{eq:secondord}
\end{equation}
where
$f(r) \equiv 1 - \frac{\omega(r)}{\Omega}$, $j \equiv e^{-\nu/2} (1 - 2GM/rc^2 )^{1/2}$, $\omega(r)$ is the rotational frequency of the local inertial frame at radius $r$, and $\Omega$ is the spin frequency of the star. The boundary conditions for the second-order partial differential equation~(\ref{eq:secondord}) are
\begin{equation}
\left[ \frac{df}{dr} \right]_{r=0} = 0
\label{eq:dfdrbc}
\end{equation}
and 
\begin{equation}
f(r = R_{\rm{NS}} ) = 1 - 2\frac{G}{c^2} \frac{I}{R^3_{\rm{NS}}}.
\label{eq:fbc}
\end{equation}
To solve these coupled equations, we integrated equations~(\ref{eq:didr}) and (\ref{eq:secondord})  outwards from the center of the star, using equation~(\ref{eq:dfdrbc}) as one boundary condition, and iterated it to find the value of $f_0$ for which equation~(\ref{eq:fbc}) is valid.

In this way, we determined the mass, radius, and moment of inertia for a given equation of state.

\section{Generating mock equations of state}
\label{sec:genmock}
In order to be as general as possible and go beyond the current sample of proposed equations of state, we tested our parametrization on a sample of mock EoS that incorporated extreme but physically allowed behavior, with the hypothesis that if our parametrization could accurately capture these extreme cases, then it would be able to reproduce more reasonable EoS as well.

\begin{figure}[h]
\centering
\includegraphics[width=0.4\textwidth]{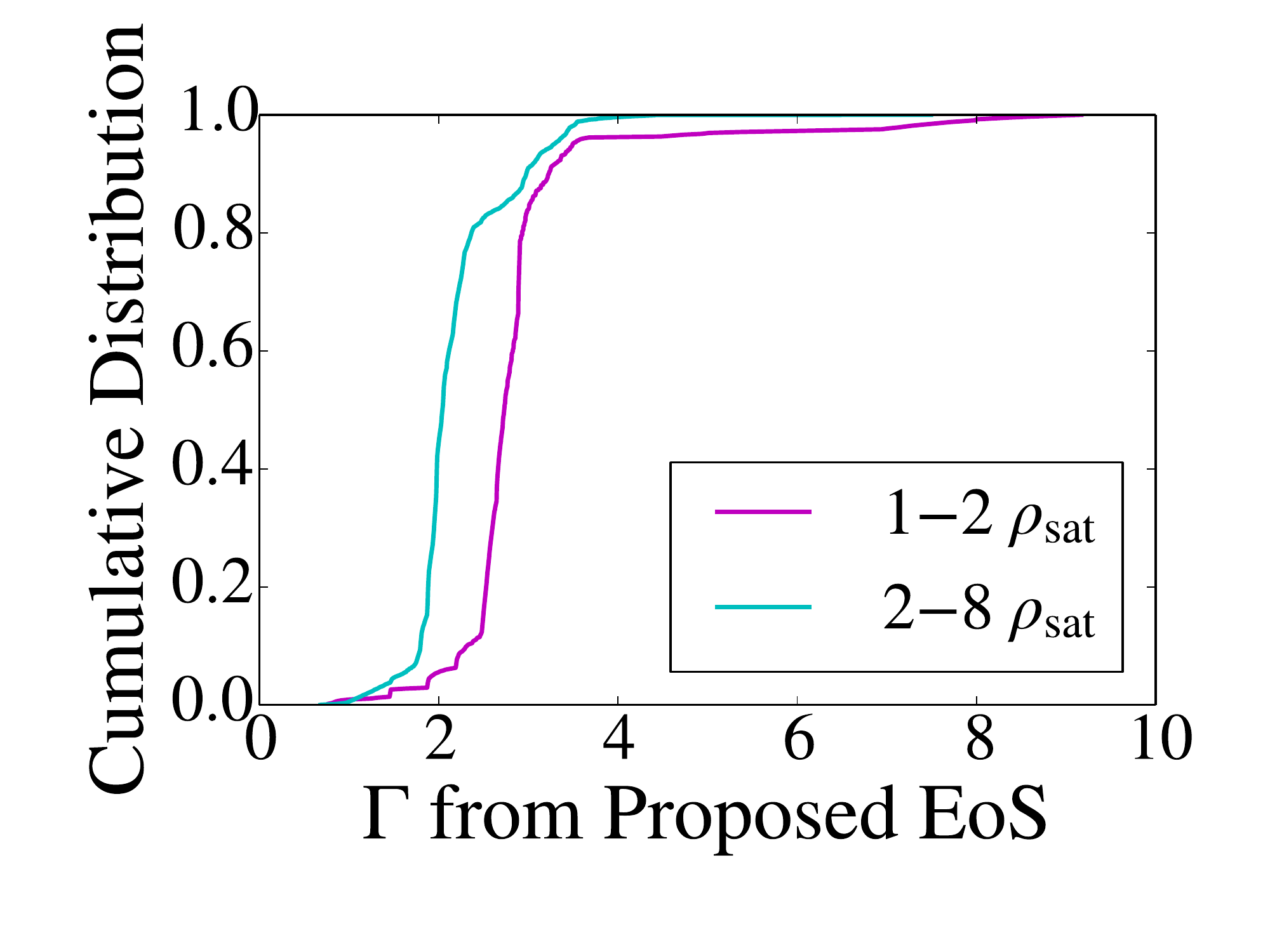}
\caption{\label{fig:gammas}  Cumulative distribution of polytropic indices, $\Gamma$, calculated at all tabulated densities in a given range in a sample of 49 proposed EoS. We found that the majority of polytropic indices lie between $\Gamma \sim$ 1 and $\Gamma \sim$ 5 and we therefore set the two extreme polytrope options in our mock EoS to have these indices.}
\end{figure}

For our mock EoS, we assumed the true EoS of neutron stars to be well known up to the nuclear saturation density. We then divided the high-density regime, from 1-8~$\rho_{\rm sat}$, into 15 segments that were evenly-spaced in the logarithm of the density, so that each segment would be small relative to the overall range of densities. The logarithmic sampling also sampled the lower density region better and thus allowed more variability in the region that most affects the global properties of the star \citep{Lattimer2001, Read2009, Ozel2009}. In each of the 15 segments, we allowed our mock equation of state to be one of two extremes, which we describe below (see $\S$~\ref{sec:phasetrans} for the addition of phase transitions to these mock EoS).

We set both the upper and lower extremes of each segment to be polytropes. To determine the polytropic index of each, we looked at a sample of 49 proposed equations of state. This sample of EoS was compiled in order to include a wide variety of physics and calculation methods, as in \citet{Read2009}. Using the tabular data for each of these EoS, we calculated the polytropic indices for all adjacent sets of pressure and density using equation~(\ref{eq:gammai}) for two density ranges. The cumulative distribution of the resulting polytropic indices is shown in Figure~\ref{fig:gammas}. We found that the majority of polytropic indices fell between $\Gamma$=1 and 5; therefore, we set our nominal upper and lower extremes to have these indices and allowed any sequence of polytropic indices across our density range.

There is, however, an absolute upper bound on the allowed polytropic index that is set by the condition of causality, which requires that the pressure gradient inside a neutron star obey the relation
\begin{equation}
\label{eq:luminal}
\frac{dP}{d\epsilon} \equiv \left(\frac{c_s}{c}\right)^2 \le 1,
\end{equation}
where $c_s$ is the local sound speed.

We can use this relation to derive the corresponding bound on $\Gamma$ by noting that we can write equation~(\ref{eq:luminal}) as
\begin{equation}
\frac{dP}{d\epsilon} = \frac{dP}{d\rho} \left( \frac{d\epsilon}{d\rho} \right)^{-1} = \left(\frac{c_s}{c}\right)^2.
\end{equation}
For a polytrope, $dP/d\rho = \Gamma P/\rho$, and the mass density-energy density relation of equation~(\ref{eq:thermo}) can be expanded to
\begin{equation}
 \frac{d\epsilon}{d\rho} = \frac{1}{\rho} \left( \epsilon + P \right).
\end{equation}
The polytropic index can therefore be written as
\begin{equation}
\label{eq:gamma_lum}
\Gamma = \frac{c_s ^2}{c^2} \frac{\epsilon + P}{P} \le \frac{\epsilon + P}{P} \equiv \Gamma_{\rm luminal}.
\end{equation}
In order to ensure that our upper extreme did not violate causality, we set our upper polytropic index to be the minimum of $\Gamma = 5$ and $\Gamma_{\rm luminal}$. 

With this set of steps, we obtained the pressure, mass density, and energy density for each segment of our mock equations of state. The mock EoS start at $\rns$, so once the pressure at this density is determined, the above relations will uniquely determine the rest of the behavior of each mock EoS. We introduced further freedom in our mock EoS by allowing two significantly different pressures at the starting point, $\rns$. This is motivated by the fact that such a bifurcation is also seen among the set of proposed EoS. We used the EoS SLy \citep{Douchin2001} and PS \citep{Pandharipande1975} to determine the lower and higher starting pressures, respectively. 

With 15 density-segments, two options for the polytropic behavior along each segment, and two options for the starting pressure, our algorithm produced $2\times2^{15}$ = 65,536 different mock EoS against which we could test our parametrization. However, we excluded any mock EoS that were too soft to produce a 1.9~$\Ms$ neutron star \citep{Demorest2010, Antoniadis2013}, reducing our final sample size to 53,343.

\subsection{The Mock EoS}
\begin{figure}[ht]
\centering
\includegraphics[width=0.39\textwidth]{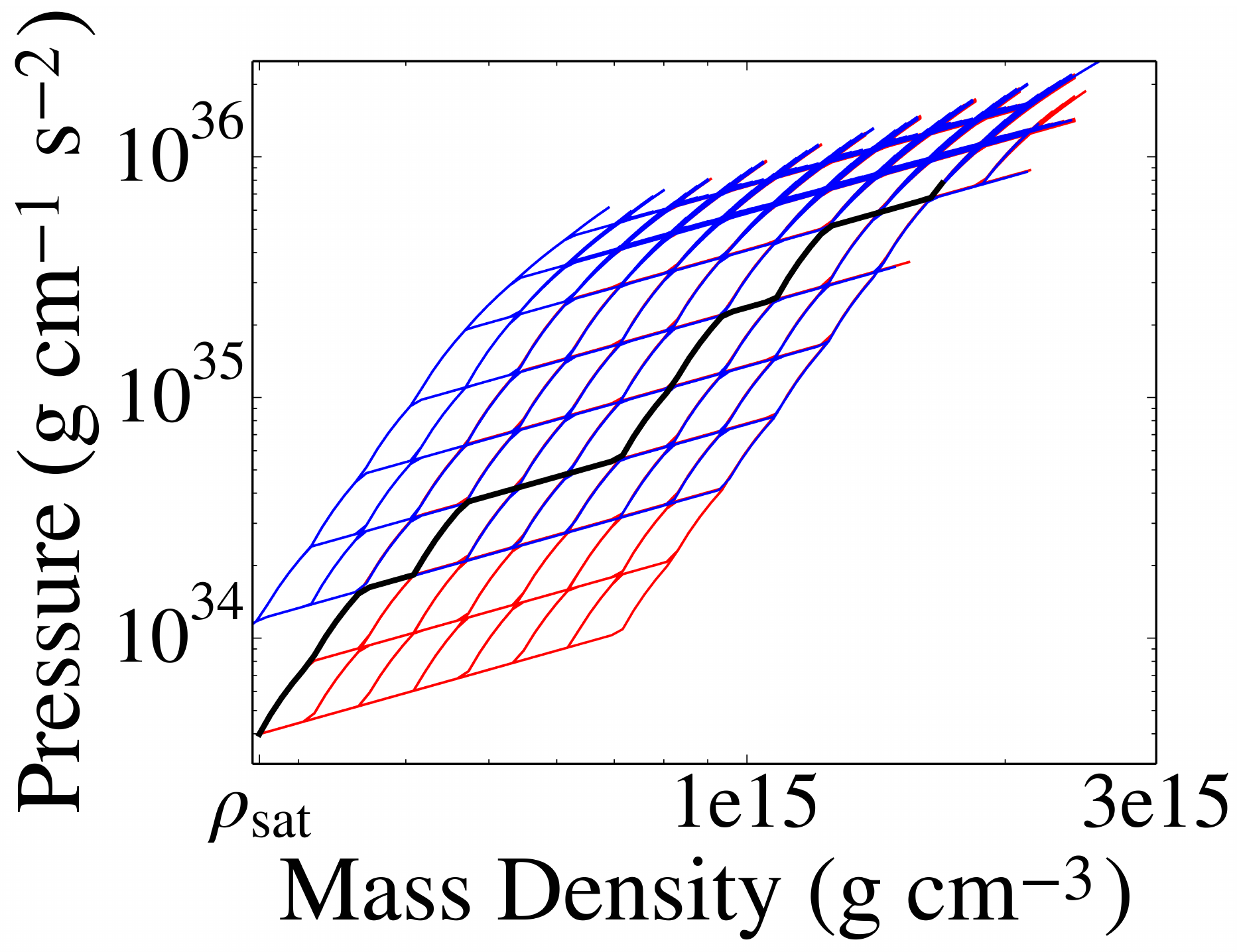}
\caption{\label{fig:mockPRho} The grid of 53,343 extreme, mock equations of state, starting at nuclear saturation density with a pressure corresponding to either the EoS SLy (in red) or PS (in blue). One sample mock EoS is shown bolded in black. Each mock EoS is composed of 15 segments that are spaced evenly in the logarithm between 1 and 8~$\rns$. Each segment is a polytrope with either $\Gamma$=1 or the minimum of $\Gamma$=5 and $\Gamma_{\rm luminal}$. Only EoS that reach 1.9~$\Ms$ are shown here, leading to the absence of segments in the lower right corner of the parameter space.}
\end{figure}

The 53,343 extreme mock equations of state that reached 1.9~$\Ms$ are shown in Figure \ref{fig:mockPRho}. 
\begin{figure}[ht]
\centering
\includegraphics[width=0.39\textwidth]{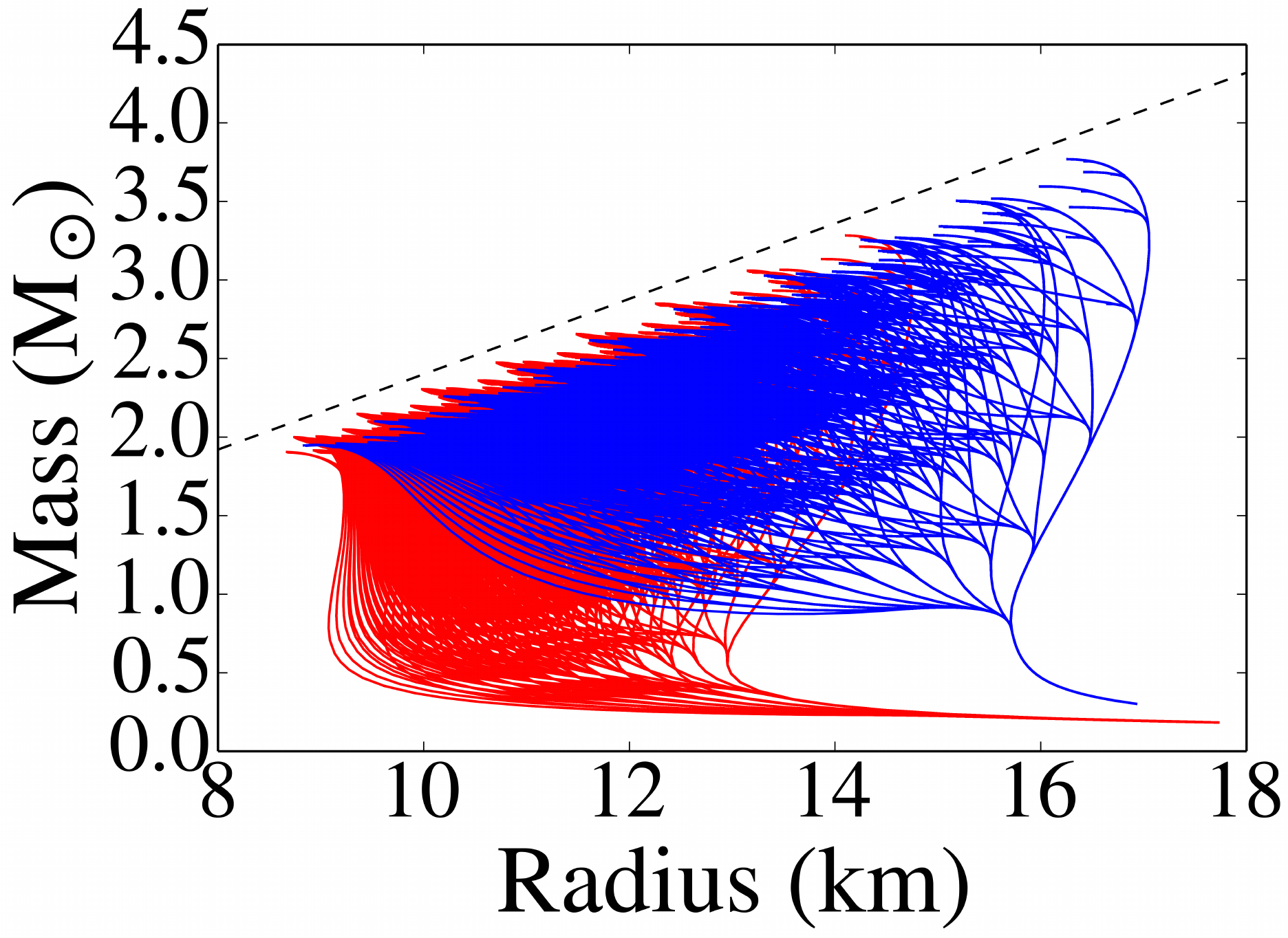}
\caption{\label{fig:mockMR} Mass-radius curves corresponding to the 53,343 mock EoS shown in Figure \ref{fig:mockPRho}. The red curves are those that started at $\rns$ with the corresponding pressure of the EoS SLy; the blue curves use the nuclear saturation pressure of EoS PS. By including both starting pressures in our sample, we are able to densely sample a realistic range of radii of $\sim$~9-17~km. These mass-radius curves also show a wide range of slopes, indicating that a large range of underlying physics is incorporated in this sample. The dashed line shows the causal relationship of $M_{\rm max} \sim 0.24 R (\Ms/ \rm km)$, derived by \citet{Lindblom1984}. The small discrepancy between this line and the observed cutoff in our M-R curves can be attributed to the different EoS that was assumed in the low-density region of the \citet{Lindblom1984} analysis.}
\end{figure}
\begin{figure*}
\includegraphics[width=\textwidth]{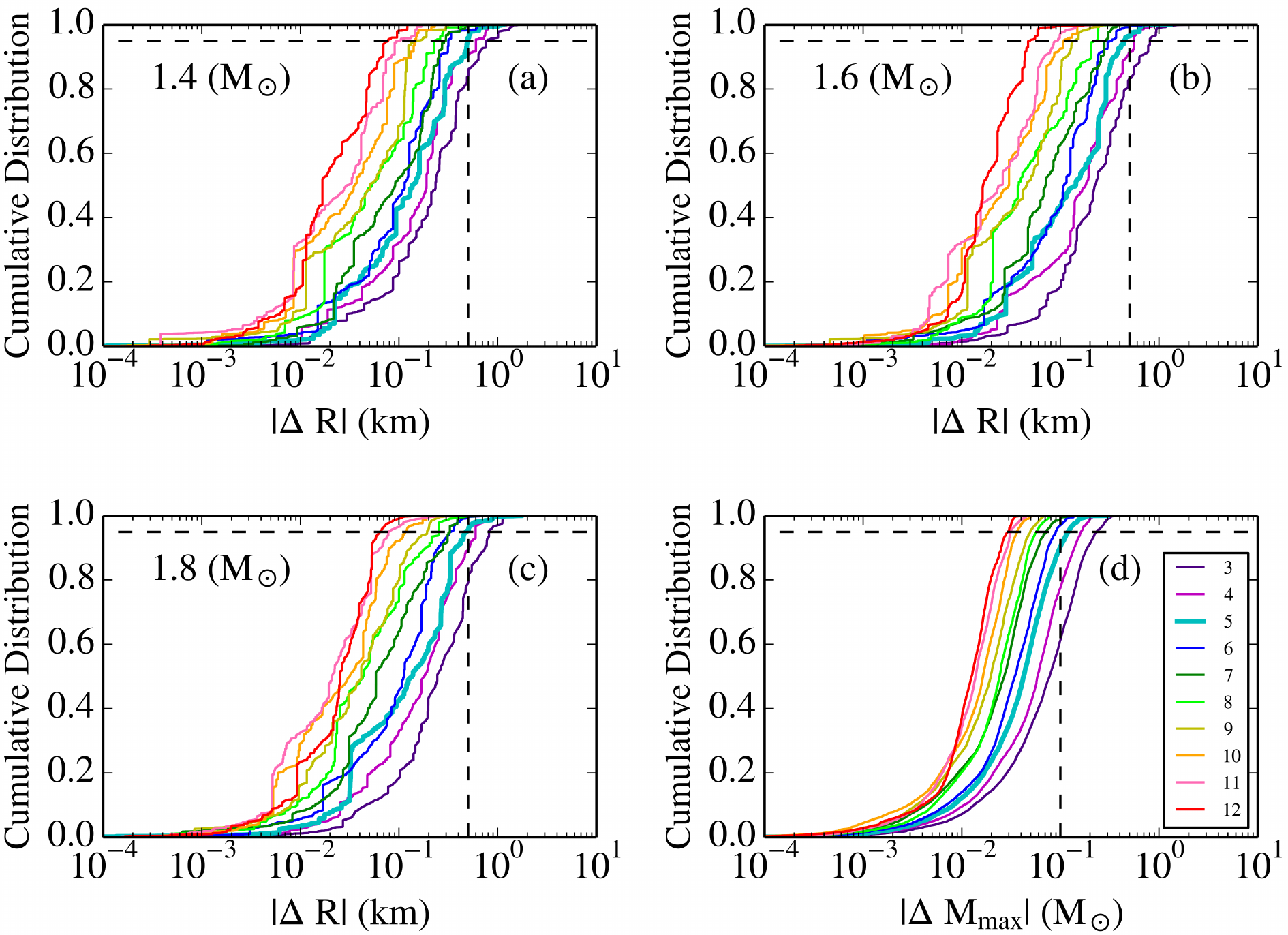}
\caption{\label{fig:deltas_noPT}  \textit{(a)-(c):} Cumulative distributions of the differences in radii between our parametrization and the full EoS for all mock EoS shown in Figure~\ref{fig:mockMR}. The different colors represent the number of fiducial densities above $\rns$ (i.e., the number of polytropic segments included in the parametrization). The radius residuals are measured at 1.4, 1.6, and 1.8~$\Ms$, respectively. The vertical dashed lines mark a residual of 0.5~km; the horizontal dashed lines mark the 95\% level of the cumulative distribution. We find that our goal of residuals $\le$0.5~km is achieved with 5 fiducial densities. \textit{(d):} Cumulative distribution of the difference in maximum mass between our parametrization and the full EoS. The lines and colors are as for the other three panels, but here the vertical dashed line is shown at 0.1~$\Ms$, corresponding to our desired maximum residual. This goal is also approximately achieved with 5 fiducial densities.}
\centering
\end{figure*}
All possible trajectories through this grid are indeed included, with the constraint that pressure must always increase with density (i.e., only monotonic behavior in this grid is allowed). The broadening of the mock EoS at high pressures is a result of our requirement that the upper polytropic limit must be the minimum of $\Gamma$=5 and $\Gamma_{\rm luminal}$. 

The corresponding mass-radius curves, calculated according to the method described in $\S$~\ref{sec:tov}, are shown in Figure~\ref{fig:mockMR}. As Figure~\ref{fig:mockMR} demonstrates, starting the mock EoS at the two different pressures corresponding to the two families of proposed EoS allowed us to fully span the range of reasonable radii. With this choice, we achieved a dense sampling of mass-radius curves that span radii from $\sim$~9-17~km for $M \lesssim 1~\Ms$. Furthermore, the mock mass-radius curves include curves that shallowly slope upwards, that are nearly vertical, and that bend backwards, indicating that we have sampled a wide range of possible underlying behavior.

The trend of increasing maximum mass with radius is a result of our causality constraint. \citet{Lindblom1984} derived the maximum gravitational redshift of a neutron star as a function of mass by assuming that the equation of state is trusted up to 3$\times 10^{14}$g~cm$^{-3}$ and configuring the resulting mass of the star to maximize the redshift. As in our analysis, that study required that the relationship between pressure and density inside the star not violate causality. After converting their relationship between the mass and maximum gravitational redshift to a mass-radius relationship, we find that the corresponding relationship of $M_{\rm max} \sim 0.24 R (\Ms/\rm km)$, shown as the dashed line in Figure~\ref{fig:mockMR}, is very close to what is seen in our mock EoS. The small differences between this relationship and that in our mock EoS can be attributed to the different EoS assumed up to the first fiducial density: \citet{Lindblom1984} assumed the EoS BPS, while we assumed either SLy or PS. Furthermore, they assumed their EoS up to $3\times10^{14}$~g~cm$^{-3}$, while we assumed SLy or PS only up to $\rns \sim 2.7\times10^{14}$~g~cm$^{-3}$.

\subsection{Determining the goodness of the parametric representation}
We quantified how well our parametrization represented the full EoS and chose the optimal number of sampling points by comparing the radii of our results to those found using the full EoS at three fiducial masses, $M$=1.4, 1.6, and 1.8~$\Ms$. We also calculated $\Delta M_{\rm max} \equiv \left|M_{\rm max}(\rm full)-\it{M}_{\rm max}(\rm parametric)\right|$ to determine how well our parametrization reproduced the maximum mass predicted by the full EoS. Finally, we calculated the difference in the moment of inertia, $\Delta I_{\rm A}$, predicted by our parametrization and by the full EoS for a star of mass $M = 1.338~\Ms$, i.e., the mass of the Pulsar A in the binary system PSR J0737$-$3039.

Given the typical uncertainties in the mass, radius, and moment of inertia, either with current data or with those expected in the near future, our goal was to reproduce the radii to within half a kilometer, i.e., to require $\Delta R <$ 0.5~km at each of the three critical masses. We also required $\Delta M_{\rm max} < 0.1~\Ms$. Because the moment of inertia for Pulsar A has not yet been measured, we did not impose strict requirements on $\Delta I_{\rm A}$ , but still included this observable in our results to see qualitatively how well it compares to some reasonable predictions for $I_{\rm A}$.

\subsection{Results of parametrization of mock EoS}
\label{sec:mockresults}
\begin{figure}
\includegraphics[width=0.5\textwidth]{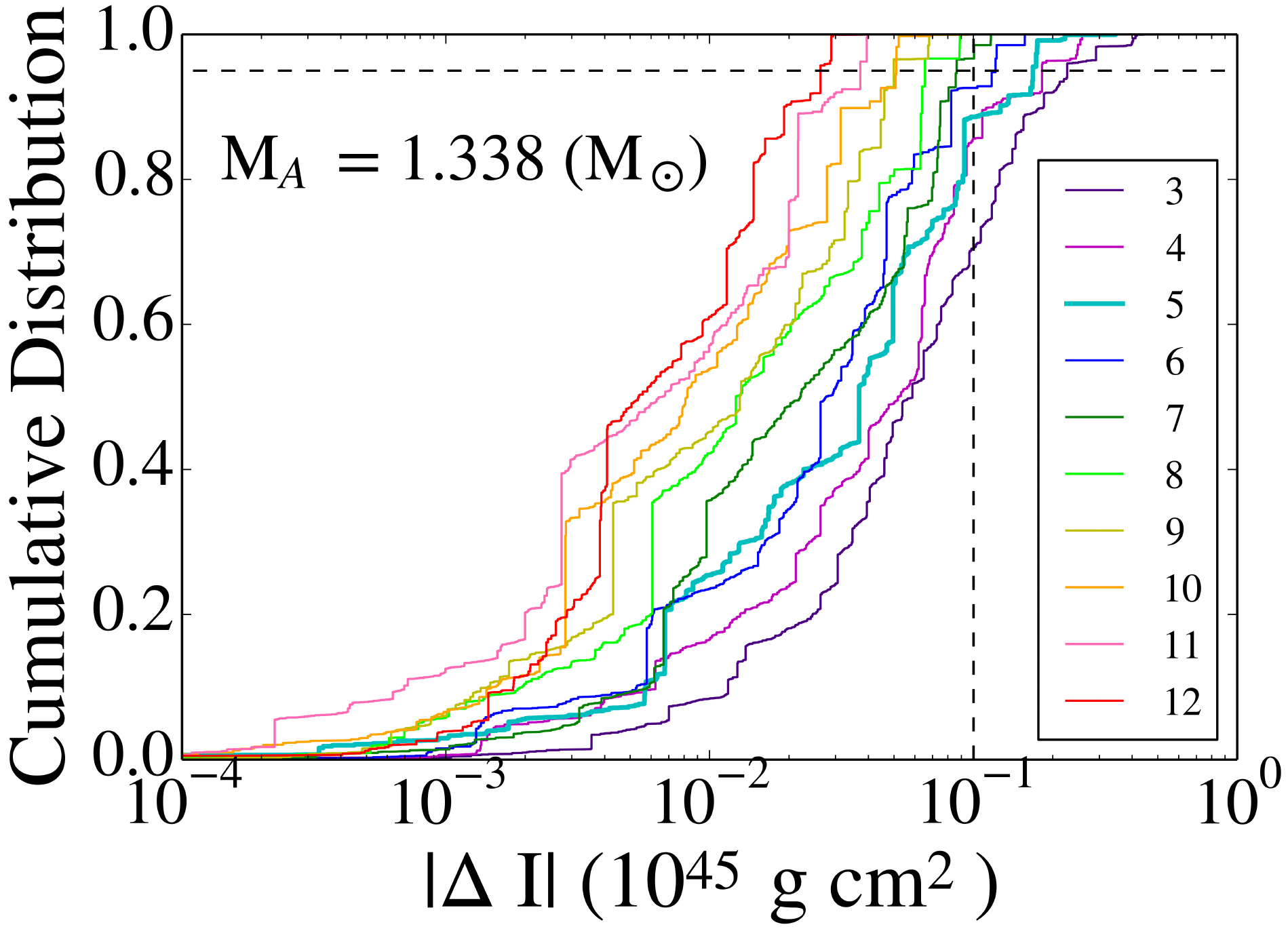}
\caption{\label{fig:deltaI_noPT} Cumulative distribution of the differences in moment of inertia for a star of mass $M_A = 1.338~\Ms$, calculated between our parametrization and the full EoS. The different colors represent the number of fiducial densities above $\rns$ that we included in our parametrization (i.e., the number of polytropic segments). The vertical dashed line marks a residual of 10\% for a hypothetical moment of inertia measurement of 10$^{45}$~g~cm$^{2}$; the horizontal dashed lines mark the 95\% level of the cumulative distribution. With 5 fiducial densities, the moment of inertia residuals are less than 0.17~$\times 10^{45}$~g~cm$^{2}$ in 95\% of cases. We therefore find that, depending on the exact value of the upcoming measurement of the moment of inertia for Pulsar A in the double pulsar system, 5-6 fiducial densities may be needed to reproduce $I_A$ to the 10\% accuracy level.  }
\centering
\end{figure}

Figure~\ref{fig:deltas_noPT} shows the  cumulative distribution of residuals in radius and mass for when various numbers of polytropic segments were included in the parametrization for our full sample of 53,343 extreme mock EoS. We found that our goal of $\Delta R < $ 0.5~km was achieved by sampling the EoS with 5 fiducial densities above $\rns$ (i.e., including 5 polytropes). Specifically, using 5 fiducial densities, we found that for 95\% of the mock EoS, $\Delta R=$ 0.50, 0.44, and 0.48 km at $M$ = 1.4, 1.6 and 1.8~$\Ms$, respectively. In addition, a parametrization with 5 fiducial densities reproduced $M_{\rm max}$ to within 0.12~$\Ms$ in 95\% of the cases. 

We also calculated the difference in the moment of inertia for a neutron star with the same mass as Pulsar A in the double pulsar system, i.e., $M_A=1.338\Ms$. The cumulative distribution of these residuals as a function of the number of polytropes included in the parametrizations is shown in Figure~\ref{fig:deltaI_noPT}. By sampling the EoS at 5 fiducial densities, we found that the residuals in moment of inertia are less than  $0.17 \times10^{45}$g~cm$^2$ in 95\% of the cases. The moment of inertia for Pulsar A is expected to be on the order of $10^{45}$g~cm$^{2}$ and is expected to be measured with 10\% accuracy \citep{Kramer2009}. As a result, depending on the exact value of the forthcoming moment of inertia measurement, 5-6 fiducial densities may be required to recreate the moment of inertia to the 10\% accuracy level.

Even a parametrization with just 3 fiducial densities reproduced the radii of $\sim$80\% of our extreme mock EoS to within 0.5~km.  However, requiring $\Delta M_{\rm max} < 0.1 \Ms$ and $\Delta~I/I \lesssim 10\%$ requires more points. We therefore conclude that, given the most recent observational uncertainties and the continuing prospects for even smaller errors in the near future, a parametrization that samples the EoS at 5 fiducial densities is optimal. Specifically, we recommend spacing the five fiducial densities evenly in the logarithm of the density, such that ($\rho_0, \rho_1, \rho_2, \rho_3, \rho_4, \rho_5) = (1.0, 1.4, 2.2, 3.3, 4.9, 7.4$)~$\rns$.

\section{Adding phase transitions to the mock EoS}
\label{sec:phasetrans}
\begin{figure*}[ht]
\includegraphics[width=\textwidth]{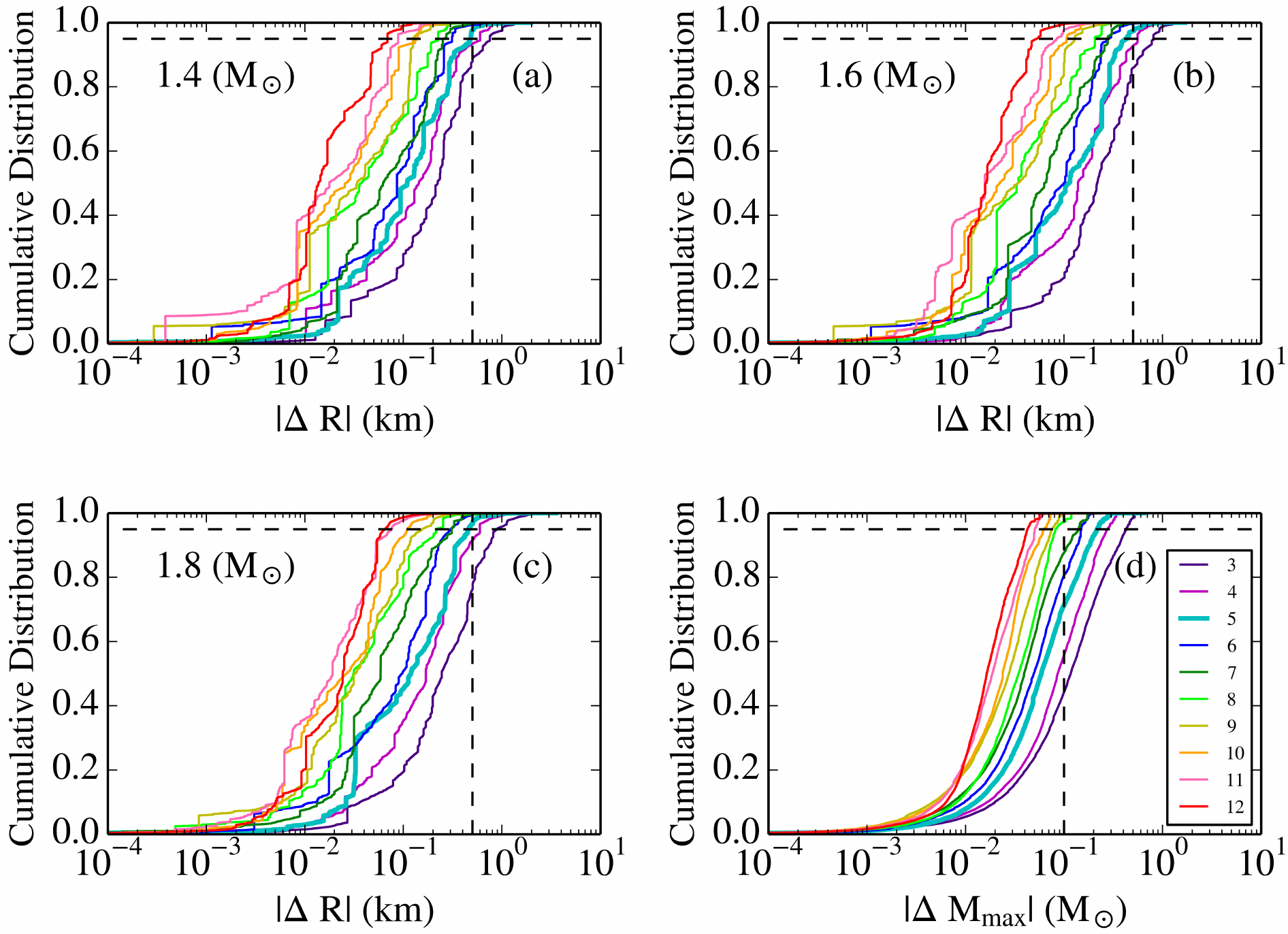}
\caption{\label{fig:deltas_PT} Same as Figure~\ref{fig:deltas_noPT} but for 176,839 randomly sampled mock EoS, each of which include a single first-order phase transition. Approximately half of this sample of mock EoS starts at a pressure corresponding to the EoS SLy at $\rns$, while the other half starts at the pressure predicted by the EoS PS at $\rns$. We find that a parametrization with five fiducial densities above $\rns$ (i.e., five polytropic segments) is sufficient to reproduce the radii at our three fiducial masses to within less than 0.5~km in 95\% of cases. The errors in maximum mass are significantly worse than for the sample of mock EoS without phase transitions: 95\% of this sample has $\Delta M_{\rm max} < 0.22 \Ms$ or less.}
\centering
\end{figure*}

We also considered more diverse equations of state by allowing there to be a first-order phase transition in the mock EoS described in $\S$~\ref{sec:genmock}. We allowed only one phase transition per mock EoS, but allowed the phase transition to start in any of our 15 density segments and to last anywhere between 1 and 15 segments. For the remaining segments, the mock EoS was polytropic with an index of $\Gamma$=1 or the minimum of $\Gamma$=5 and $\Gamma_{\rm luminal}$, as above. With the addition of these phase transitions, there are now $2^{N} \times N(N+1) / 2$ possibilities for $N$ segments of the EoS, for each possible starting pressure. For 15 segments and our two starting pressures (corresponding to the EoS SLy and PS at $\rns$), this corresponds to 7,864,320 possibilities for the mock EoS.

We randomly sampled $\sim$175,000 of these mock EoS, with roughly half starting at each of the initial pressures. We then applied our parametrization to each mock EoS. As we did for the original set of mock EoS, we varied the number of polytropic segments in the parametrization between 3 and 12 and calculated the resulting residuals in radius, mass, and moment of inertia. The mass and radius residuals are shown in Figure~\ref{fig:deltas_PT}. With 5 fiducial densities above $\rns$, i.e., the optimal number of polytropic segments found in $\S$~\ref{sec:mockresults}, we found that 95\% of the radius residuals were less than 0.48, 0.43, and 0.46~km at 1.4, 1.6, and 1.8~$\Ms$, respectively. These errors are comparable to those from the mock EoS without phase transitions. We also found that 95\% of the differences in maximum mass were less than 0.22~$\Ms$, which is a larger error than in our previous, less extreme sample of mock EoS. However, 70\% of the errors in maximum mass were still less than 0.1~$\Ms$ for this sample, indicating that this parametrization reasonably recreated the maximum mass for many of our most extreme sample of mock EoS. The error distribution for the moment of inertia was almost identical to the distribution for the sample without phase transitions. We found that a parametrization with 5 fiducial densities reproduced the moment of inertia to within 0.17~$\times 10^{45}$~g~cm$^{2}$ for 95\% of the mock EoS with phase transitions.

The tail end of all four distributions in Figure~\ref{fig:deltas_PT} extended to higher errors than did the tail in Figure~\ref{fig:deltas_noPT}. This is because the arbitrarily-placed phase transitions in this sample of mock EoS can make the resulting mass-radius curve have very sharp turn-overs. If a sharp turn-over occurs near one of the fiducial masses at which we measure radius residuals, we will infer an artificially high error. However, the distributions in Figure~\ref{fig:deltas_PT} show that the probability of the true EoS falling in this tail is small ($\lesssim~5\%$ with 5 fiducial densities). We therefore find that, for the vast majority of cases, a parametrization that samples the EoS at 5 fiducial densities is sufficient to reproduce radius and maximum mass observables to within 0.5~km and 0.1-0.2~$\Ms$, even if the EoS contains a first-order phase transition at an arbitrary density and over an arbitrary range.

\section{Application of the parametrization to physically-motivated EoS}
\label{sec:tabEoS}
\begin{figure*}
\includegraphics[width=\textwidth]{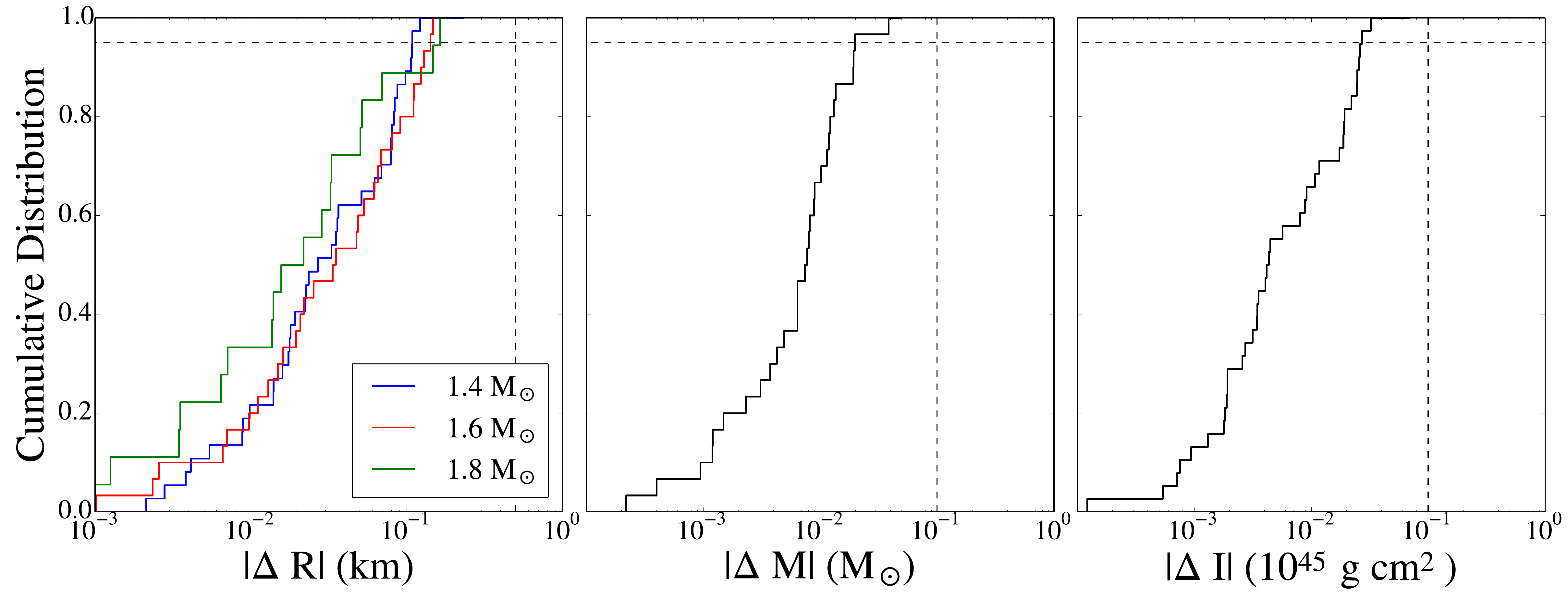}
\caption{\label{fig:deltaRMI_tab} Cumulative distribution of the residuals measured between the full EoS and the parametric version for 42 proposed EoS. The parametrization uses 5 fiducial densities above $\rns$ (i.e., it includes 5 polytropic segments). \textit{Left panel: } Residuals in radius, as calculated at 1.4, 1.6, and 1.8 $\Ms$. The vertical dashed line indicates residuals of 0.5~km, while the horizontal line shows the 95\% inclusion level (this 95\% inclusion line is identical in all three panels). We find that the residuals are less than 0.10, 0.12, and 0.09~km at 1.4, 1.6, and 1.8~$\Ms$ respectively for 95\% of the proposed EoS. \textit{Middle panel: } Differences in maximum mass. The vertical dashed line indicates residuals of 0.1~$\Ms$. We find that the errors in maximum mass are less than 0.04~$\Ms$ for 95\% of the proposed EoS. \textit{Right panel: } Differences in the moment of inertia for a star of mass $M_A = 1.338 \Ms$. The vertical dashed line indicates residuals of 10\% for a hypothetical moment of inertia measurement of 10$^{45}$~g~cm$^{2}$. We find that 95\% of the proposed EoS have residuals in the moment of inertia of $0.02 \times 10^{45}$~g~cm$^{2}$ or smaller. }
\centering
\end{figure*}
Even though we optimized our parametrization using EoS that span a much wider range of possibilities than the currently proposed ones, we nevertheless explored how well this parametrization reproduced the physically-motivated EoS found in the literature. To this end, we applied our optimized parametrization (5 fiducial densities above $\rns$) to a sample of 42 proposed EoS,\footnote{This sample is smaller than the 49 EoS that we previously cited because we exclude from this subsample any calculated EoS that become acausal, except for the EoS AP4. Even though AP4 reaches a local sound speed of $c_s \sim 1.1c$ by densities of $\rho \sim6~\rns$ and becomes more acausal thereafter, we do include this EoS, as it is commonly used and included in the literature. We also exclude from this sample two EoS that are not calculated to high enough densities to accommodate our parametrization at 7.4~$\rns$.} which incorporate a variety of different physical possibilities and calculation methods. (The tabular data for these EoS are compiled in \citealt{Cook1994, Lattimer2001, Read2009}; and \citealt{Ozel2016}).  Our sample included purely nucleonic equations of state, such as: relativistic (BPAL12 and ENG) and nonrelativistic (BBB2) Brueckner-Hartree-Fock EoS; variational-method EoS (e.g. FPS and WFF3); and a potential-method EoS (SLy). We also include models which incorporate more exotic particles, including, for example, a neutron-only EoS with pion condensates (PS), a relativistic mean-field theory EoS with hyperons and quarks (PCL2), and an effective-potential EoS with hyperons (BGN1H1). 

In applying our parametrization to these proposed EoS, we no longer connected to SLy or PS for $\rho < \rns$. Instead, we assumed that each EoS is known up to $\rns$, and we therefore used the full EoS that we were parametrizing for the low-density regime. For $\rho > \rns$, we applied our parametrization as above. After applying our parametrization, we calculated the resulting residuals in radii at the three fiducial masses,\footnote{For every EoS, we only calculated and included the radius resdiual at a given mass if the EoS actually reaches that mass. If, for example, an EoS only produces masses up to 1.7~$\Ms$, we still included the radius residuals at 1.4 and 1.6~$\Ms$ and simply excluded the data point at 1.8~$\Ms$.} the maximum mass, and the moment of inertia. The residuals for our optimized, 5-polytrope parametrization are shown in Figure~\ref{fig:deltaRMI_tab}.

\begin{figure}[ht]
\includegraphics[width=0.5\textwidth]{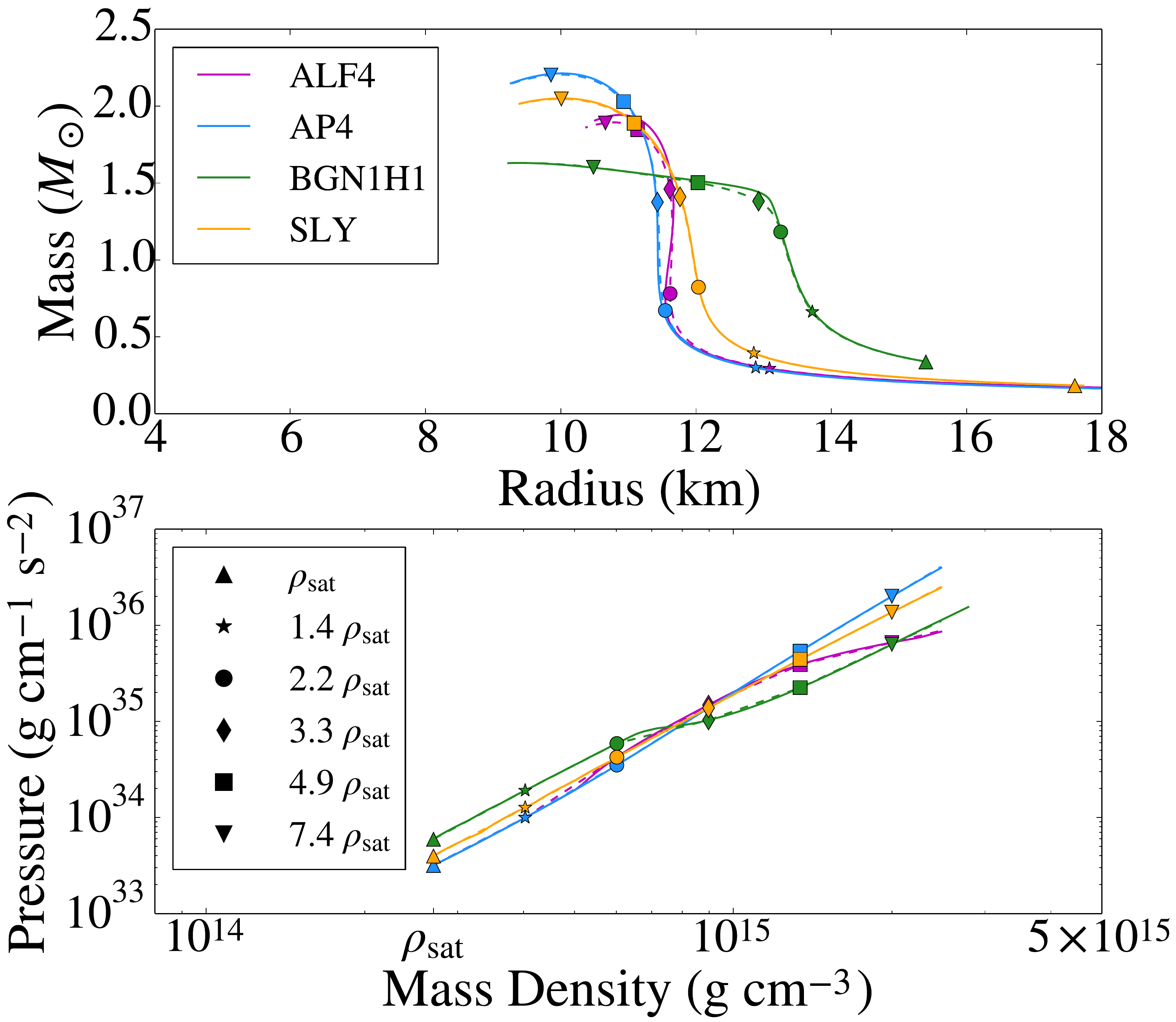}
\caption{\label{fig:ap4} \textit{Top: } Mass-radius relations for the EoS AP4, AL4, BGN1H1, and SLy as solid lines. The dashed lines show our parametrization of each, with five fiducial densities above $\rns$. The different symbols represent the mass and radius of a star with a central density equal to each fiducial density. \textit{Bottom: }  Pressure as a function of mass density for these EoS. The symbols represent the location of each fiducial density. We find that this parametrization reproduces the EoS to very high accuracy in mass-radius space.}
\centering
\end{figure}

The errors in applying the parametrization to the proposed, physically motivated EoS were much lower than for the more extreme, mock EoS. Our optimized parametrization reproduced the radii of $\sim$95\% of the proposed EoS to within 0.10, 0.12, and 0.09~km (at 1.4, 1.6, and 1.8~$\Ms$, respectively) and the maximum masses of 95\% of the EoS to within 0.04 $\Ms$.  As examples, we show in Figure~\ref{fig:ap4} the full mass-radius relation as well as the one calculated from our parametrized EoS for several proposed EoS: AP4 (nucleonic), ALF4 (quark hybrid), SLy (nucleonic), and BGN1H1 (includes hyperons).  As seen here, the differences in mass-radius space are extremely small.

\section{Conclusions}
In this paper, we investigated an optimal parametrization of the neutron star equation of state that can be used to interpret neutron star observations. We found that a parametric EoS with five polytropic segments evenly spaced in logarithm between 1 and 7.4~$\rns$ was sufficient to reproduce the radii of proposed EoS to within 0.12~km and the maximum mass to within 0.04~$\Ms$ in 95\% of cases. This parametrization was also able to reproduce the radii of our more extreme, mock EoS to within 0.5~km, suggesting that even if a more extreme EoS is proposed or realized in nature, our parametrization will be robust enough to reproduce it well.

The radii of approximately fifteen neutron stars have already been measured, for most of which the masses are also known \citep{Guillot2013, Guillot2014, Heinke2014, Nattila2015,Ozel2016a, Bogdanov2016}. Even though the uncertainties in the individual radius measurements are of order $\sim$2~km, combining all the measurements leads to a rather narrow range of predicted neutron star radii. Moreover, these spectroscopic measurements will be independently checked in the near future by, e.g., waveform modeling with NICER \citep{Gendreau2012}. The forthcoming measurement of the moment of inertia, which is a higher-order moment of the mass distribution of the star \citep{Lyne2004, Kramer2009}, and the prospect of eventual measurements of other higher-order moments using gravitational wave observations from coalescing neutron stars (e.g., \citealt{Read2009b, Del-Pozzo2013, Agathos2015}) will provide additional constraints. Finally, the recent mass mesaurements of approximately two-solar mass neutron stars \citep{Demorest2010, Antoniadis2013} already place strong priors on empirically inferred EoS. 

Previous empirical inferences of the equation of state based on existing observations typically relied on functional forms with 3-4 parameters (e.g., \citealt{Ozel2010, Steiner2010,Ozel2016a, Steiner2016}). It is difficult to compare the errors produced in these parametrizations directly to our own, as they are not always quantified in mass-radius space. \citet{Read2009} report errors in their parametrization in pressure vs. density:  they find that a four-parameter EoS produced average errors in $\log P$ of 0.013, which corresponds to a fractional error in pressure of $\sim$3\%. Similarly, \citet{Lindblom2012} report errors in their spectral parametrization in energy density vs. enthalpy space:  their errors ranged from 1 to 15\% when they used 4$-$5 spectral parameters. Finally, \citet{Steiner2013} performed a preliminary comparison between five different types of models (e.g., a parametrization of two polytropes and one with four line segments) and found the largest difference in the predicted radii between any two of their models to be 0.8~km; however, these models were not necessarily optimized in terms of the number of parameters.

Ultimately, it is the inverse process of what we have shown that will be of interest: the
inference of a parametric EoS from astrophysical measurements. We will address this
aspect of the problem in a forthcoming paper. However, our work here suggests that the
uncertainties in the inference of the EoS will be dominated by the quality of the data
and the uncertainties in the inversion process itself, rather than those introduced by
the parametrization of the EoS.
\\ \\
{\em{Acknowledgements.\/}} We thank Gordon Baym for useful comments on the manuscript. We gratefully acknowledge support from NASA grant NNX16AC56G. 

\bibliography{carolyn_2}
\bibliographystyle{apj}
\FloatBarrier

\appendix
\renewcommand{\theequation}{A\arabic{equation}}

For completeness, we also explored a parametrization that uses linear segments between a number of density points to represent the EoS. As in the case of our polytropic parametrization, we started this parameterization at $\rns$ and spaced the segments evenly in the logarithm of the density. The EoS along each linear segment is given by
\begin{equation}
P = m_i \rho + b_i  \quad\quad (\rho_{i-1} \le \rho \le \rho_i),
\end{equation} 
where continuity at the endpoints implies
\begin{equation}
m_i = \frac{P_i - P_{i-1}}{\rho_i - \rho_{i-1}}
\end{equation}
and
\begin{equation}
b_i = P_{i-1} - \left(\frac{P_i - P_{i-1}}{\rho_i - \rho_{i-1}}\right) \rho_{i-1}.
\end{equation}

Using this linear relationship for pressure to integrate the differential equation~(\ref{eq:thermo}), we find that the energy density in this case is given by
\begin{equation}
\epsilon(\rho) = (1 + a)\rho c^2  + m \rho \log{\rho} - b,
\end{equation}
where $a$ is an integration constant. By requiring continuity in the energy density at the endpoints of each segment, we can solve for the integration constant such that
\begin{equation}
\label{eq:lineareps}
\epsilon(\rho) =  \left( \frac{\epsilon_{i-1} + b_i}{\rho_{i-1}} - m_i \log{\rho_{i-1}}\right) \rho + m_i \rho \log{\rho} - b_i,   \quad (\rho_{i-1} \le \rho \le \rho_i) .
\end{equation}
We used equation~(\ref{eq:lineareps}) to relate the linear EoS to the energy density and then used that energy density to integrate the TOV equations and solve for the total mass, radius, and moment of inertia of the neutron star.

\begin{figure*}
\includegraphics[width=\textwidth]{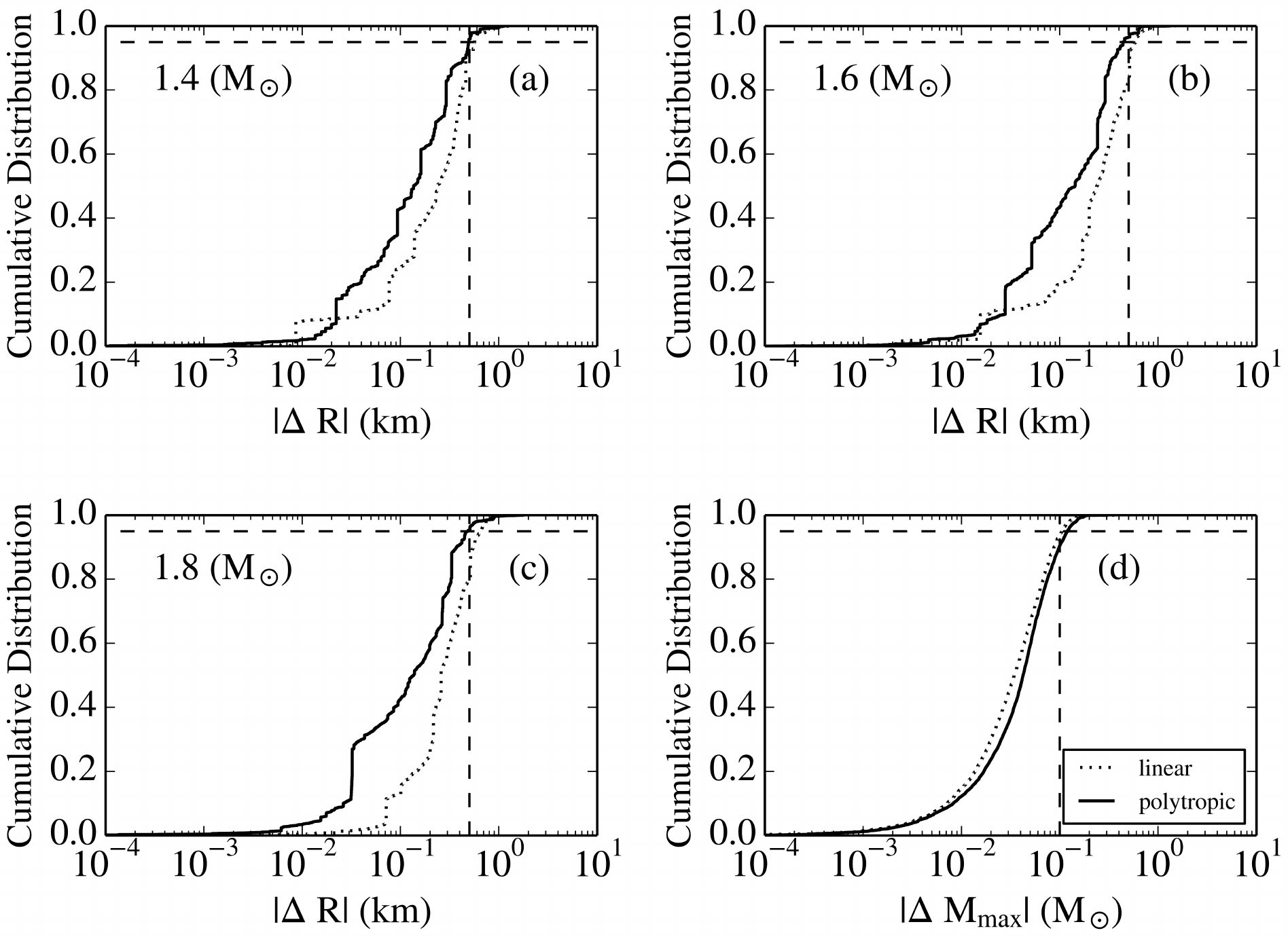}
\caption{\label{fig:deltas_linear}  \textit{(a)-(c):} Cumulative distributions of the differences in radii between a polytropic parametrization and the full EoS (solid line) and the differences between a linear parametrization and the full EoS (dashed line). These differences were calculated for all mock EoS shown in Figure~\ref{fig:mockMR}. Both parametrizations contain five fiducial densities (i.e., five segments). The radius residuals are measured at 1.4, 1.6, and 1.8~$\Ms$, respectively. The vertical dashed lines mark a residual of 0.5~km; the horizontal dashed lines mark the 95\% level of the cumulative distribution.  We find that a polytropic parametrization results in smaller errors, but that the linear parametrization still achieves the desired radius residual of 0.5~km for $\sim$80\% of our extreme, mock EoS.  \textit{(d):} Cumulative distribution of the difference in maximum mass between each parametrization and the full EoS. The lines and linestyles are as for the other three panels, but here the vertical dashed line is shown at 0.1~$\Ms$, corresponding to our desired maximum residual. We find that the linear and polytropic parametrizations perform comparably well in recreating the neutron star maximum mass.}
\centering
\end{figure*}

We applied a five-segment parametrization to the $\sim$53,000 mock EoS shown in Figure~\ref{fig:mockMR} and calculated the errors in radius at 1.4, 1.6, and 1.8~$\Ms$ as well as the errors in the maximum mass. In this way, we can directly compare these results with those from the five-polytrope optimal parametrization in Figure~\ref{fig:deltas_linear}. We find that the polytropic parametrization performs modestly better. However, the linear parametrization is still able to recreate the radii of the full EoS to within $\lesssim 0.5$~km for $\sim$80\% of the extreme, mock EoS.

Even though our mock EoS are composed of polytropic segments, there are multiple mock EoS segments per parametrization segment. Moreover, the mock EoS segments are offset from the parametrization segments. We therefore do not expect that the mock EoS should significantly bias the performance of a polytropic parametrization over a linear parametrization.

\begin{figure*}
\includegraphics[width=\textwidth]{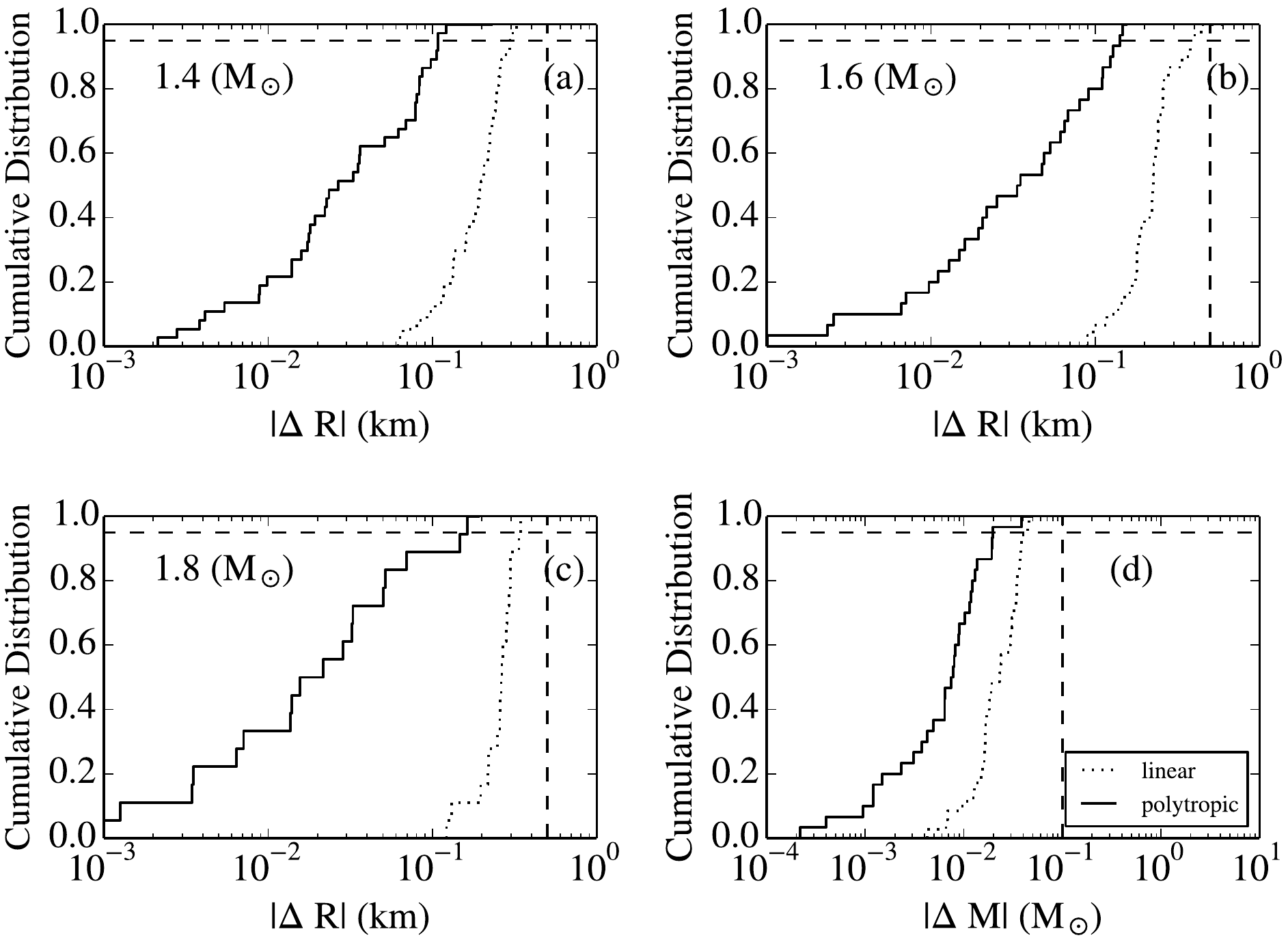}
\caption{\label{fig:deltas_linear_tab}  Same as Figure~\ref{fig:deltas_linear} but for 42 proposed, physically-motivated EoS. We find that a polytropic parametrization results in smaller errors, but that both parametrizations are able to recreate the radii and maximum masses of the full EoS to well below the expected observational uncertainties of 0.5~km and 0.1~$\Ms$, respectively.}
\centering
\end{figure*}

In order to see how well the linear parametrization performs for more physically motivated EoS, we also applied it to the sample of 42 proposed EoS from $\S$~\ref{sec:tabEoS}. We compare these results to those of the polytropic parametrization in Figure~\ref{fig:deltas_linear_tab}. We find again that the polytropic parametrization performs better than the linear parametrization, although the differences between the two parametrizations are most significant at radius and maximum mass errors that are well below observational uncertainties.

It is likely that the same levels of errors could be achieved by the linear parametrization if more than five segments were included; however, for five segments, the polytropic parametrization performs modestly better in most cases. A polytropic parametrization is also the more natural choice for the neutron star EoS. We, therefore, recommend a five-segment polytropic parametrization over a linear one.


\end{document}